\documentclass[twocolumn,showpacs,preprintnumbers,amsmath,amssymb]{revtex4}
\usepackage{graphicx}
\usepackage{dcolumn}
\usepackage{bm}
\usepackage{epsfig}
\newcommand{\be}{\begin{equation}}
\newcommand{\ee}{\end{equation}}
\newcommand{\ba}{\begin{eqnarray}}
\newcommand{\ea}{\end{eqnarray}}
\newcommand{\bi}{\begin{itemize}}
\newcommand{\ei}{\end{itemize}}
\newcommand{\tr}{{\rm Tr\,}}
\newcommand{\re}{\mathop{\rm Re}}

\newcommand{\nn}{\nonumber \\}

\newcommand{\ovl}{\overline}

\newcommand{\half}{{\textstyle\frac{1}{2}}}

\newcommand{\quarter}{{\textstyle\frac{1}{4}}}

\newcommand{\<}{\langle}
\renewcommand{\>}{\rangle}
\newcommand{\eq}{Eq.~}

\newcommand{\fig}{Fig.~}

\newcommand{\la}{\label}

\begin{document}

\preprint{MIT-CTP 3830}

\title{A calculation of the shear viscosity in SU(3) gluodynamics} 

\author{Harvey~B.~Meyer}
\email{meyerh@mit.edu}
\affiliation{Center for Theoretical Physics\\ Massachusetts Institute of Technology\\
Cambridge, MA 02139, U.S.A.}

\date{\today}

\begin{abstract}
We perform a lattice Monte-Carlo calculation of the two-point functions
of the energy-momentum tensor at finite temperature in the SU(3) gauge theory.
Unprecedented precision is obtained thanks to a multi-level algorithm. The lattice 
operators are renormalized non-perturbatively and the classical discretization 
errors affecting the correlators are corrected for. A robust upper bound for 
the shear viscosity to entropy density ratio is derived, $\eta/s < 1.0$,
and our best estimate is $\eta/s = 0.134(33)$ at $T=1.65T_c$ 
under the  assumption of smoothness of the spectral function in the low-frequency region.
\end{abstract}

\maketitle

\noindent\textit{Introduction.---~~}
Models treating the system produced in heavy ion collisions at RHIC as an ideal fluid
have had significant success in describing the observed flow phenomena~\cite{huovinen,shuryak}. 
Subsequently the leading corrections due to a finite shear viscosity were computed~\cite{teaney},
in particular the flattening of the elliptic flow coefficient 
$v_2(p_{\rm T})$ above 1GeV.
It is therefore important to compute the QCD shear and bulk viscosities from first principles
to  establish  this description more firmly.
Small transport coefficients are a signature of strong interactions, which lead
to efficient transmission of momentum in the system. Strong interactions in turn 
require non-perturbative computational techniques. 
Several attempts have been made to compute these observables
on the lattice in the SU(3) gauge theory~\cite{karsch-visco,nakamura}. 
The underlying basis of these calculations
are the Kubo formulas, which relate each transport coefficient to a spectral function $\rho(\omega)$
at vanishing frequency. Even on current computers, these calculations
are highly non-trivial, due to the  fall-off of the relevant correlators in Euclidean time
(as $x_0^{-5}$ at short distances), implying a poor signal-to-noise ratio in a standard 
Monte-Carlo calculation. The second difficulty is to solve the ill-posed inverse problem
for $\rho(\omega)$ given the Euclidean correlator at a finite set of points.
Mathematically speaking, the uncertainty on a transport coefficient $\chi$
is infinite for any finite statistical accuracy, because 
adding $\epsilon \omega\delta(\omega)$ to $\rho(\omega)$ merely corresponds to adding 
a constant to the Euclidean correlator of order $\epsilon$, while
rendering $\chi$ infinite. Therefore smoothness assumptions 
on $\rho(\omega)$ have to be made, which are reasonable far from the one-particle energy
eigenstates, and can be proved in the hard-thermal-loop framework~\cite{aarts}.

In this Letter we present a new calculation which dramatically improves on the 
statistical accuracy of the Euclidean correlator relevant to the shear viscosity
through the use of a two-level algorithm~\cite{hm-ymills}.
This allows us to derive a robust upper bound on the viscosity and a useful
estimate of the ratio $\eta/s$, which has acquired a special significance since
its value $1/4\pi$ in a class of strongly coupled supersymmetric 
gauge theories~\cite{policastro}
was conjectured to be an absolute lower bound
for all substances~\cite{kovtun}.

\noindent\textit{Methodology.---~~}
In the continuum, the energy-momentum tensor $\ovl T_{\mu\nu}(x)=F_{\mu\alpha}^a F_{\nu\alpha}^a 
             -\quarter \delta_{\mu\nu}  F_{\rho\sigma}^a F_{\rho\sigma}^a$, 
being a set of Noether currents associated with 
translations in space and time, does not renormalize.
With $L_0=1/T$ the inverse temperature, we consider the Euclidean two-point function
($0<x_0<L_0$)
\be
C(x_0)  =  L_0^5\int d^3{\bf x} ~\<\ovl T_{12}(0)\ovl T_{12}(x_0,{\bf x}) \>.
\la{eq:Cx0}
\ee
The  tree-level expression is
 $C^{\rm t.l.}(x_0) = \frac{32d_A}{5\pi^2} \Big(f(\tau)-\frac{\pi^4}{72} \Big)$,
with $\tau=1-\frac{2x_0}{L_0}$, $d_A=8$ the number of gluons 
and $f(z) = \int_0^\infty ds~ s^4  \cosh^2(z s)/\sinh^2 s$.
The correlator $C(x_0)$ is thus dimensionless and, in a conformal field theory, 
would be a function of $Tx_0$ only. 

The spectral function is defined by 
\be
C(x_0) = L_0^5
\int_0^\infty \rho(\omega) \frac{\cosh \omega(\half L_0-x_0)}{\sinh \frac{\omega L_0}{2}} d\omega.
\la{eq:C=int_rho}
\ee
The shear viscosity is given by~\cite{hosoya,karsch-visco}
\be
\eta(T) = \pi \left.\frac{d\rho}{d\omega}\right|_{\omega=0}.
\ee
Important properties of $\rho$ are its positivity, $\rho(\omega)/\omega\geq0$
and parity, $\rho(-\omega)=-\rho(\omega)$.
The spectral function that reproduces $C^{\rm t.l.}(x_0)$ is
\ba
\rho^{\rm t.l.}(\omega) &=& 
 \frac{A_{\rm t.l.} ~\omega^4 }{\tanh\quarter \omega L_0} 
     + BL_0^{-4}~\omega \delta(\omega) , \la{eq:tl}\\
A_{\rm t.l.} &=& \frac{1}{10} \frac{d_A}{(4\pi)^2},~~~~~
B = \left(\frac{2\pi}{15}\right)^2 ~ d_A .
\ea
While the $\omega^4$ term is expected to survive in the interacting theory with only
logarithmic corrections, the $\delta$-function at the origin corresponds to the fact 
that gluons are asymptotic states in the free theory and implies an infinite viscosity.

On the lattice, translations only form a discrete group, so that a finite renormalization 
is necessary,
$\overline T_{\mu\nu}(g_0) = Z(g_0) \overline T_{\mu\nu}^{\rm (bare)}.$
We employ the Wilson action~\cite{wilson74},
$ S_{\rm g} =  \frac{1}{g_0^2} \sum_{x,\mu\neq\nu} \tr\{1-P_{\mu\nu}(x)\}$,
on an $L_0\cdot L^3$  hypertoroidal lattice,
and the following discretized expression of the Euclidean  energy:
\[ \overline T_{00}^{\rm (bare)}(x) 
\equiv \frac{2}{a^4g_0^2} \Big[ \sum_{k<l} \re\tr P_{kl}(x) - \sum_{k} \re\tr P_{0k}(x)  \Big]
\]
One of the lattice sum rules~\cite{michael2} can be interpreted as a non-perturbative 
renormalization condition for this particular discretization, from which 
we read off  $ Z(g_0)  = 1- \half g_0^2  (c_\sigma-c_\tau)$.
The definition of the anisotropy coefficients $c_{\sigma,\tau}$ can be found
in~\cite{karsch-aniso}, where they are computed non-perturbatively.
With a precision of about $1\%$, a Pad\'e fit constrained by 
the one-loop result~\cite{karsch-pert} yields
\be
Z(g_0) = \frac{1 -1.0225 g_0^2 + 0.1305 g_0^4}{ 1- 0.8557g_0^2},\quad (6/g_0^2\geq5.7).
\ee

\noindent\textit{Numerical results.---~~}
We report results obtained on a $\beta\equiv6/g_0^2=6.2$, $8\cdot20^3$ lattice
and on a $\beta=6.408$, $8\cdot28^3$ lattice. The first is thus at a temperature of 
$1.24T_c$, the second at $T=1.65T_c$. We use the results for $aT_c$ obtained 
in~\cite{teper-sun} and the non-perturbative lattice $\beta$-function
of~\cite{necco-sommer} to determine this.
We employ the two-level algorithm described in~\cite{hm-ymills}. The computing time
invested into the $1.65T_c$ simulation is about 860 PC days.
Following~\cite{karsch-visco}, 
we discretize $\quarter\<(\ovl T_{11}-\ovl T_{22})(\ovl T_{11}-\ovl T_{22})\>$ 
instead of $\<\ovl T_{12}\ovl T_{12}\>$ (the two are equal in the continuum)
to write 
$C(x_0)=\frac{L_0^5}{L^3}\< O_\eta(0)O_\eta(x_0)\>+{\rm O}(a^2)$, where
\ba
&&O_\eta(x_0) \equiv \half a^3 \sum_{\bf x} \{\ovl T_{11} -\ovl T_{22} \}(g_0,x)\nn
& &=\frac{2Z(g_0)}{a~g_0^2}~ \sum_{\bf x} \re\tr\{P_{10}+P_{13}-P_{20}-P_{23}\}(x).\nonumber
\ea

\begin{figure}
\begin{center}
\psfig{file=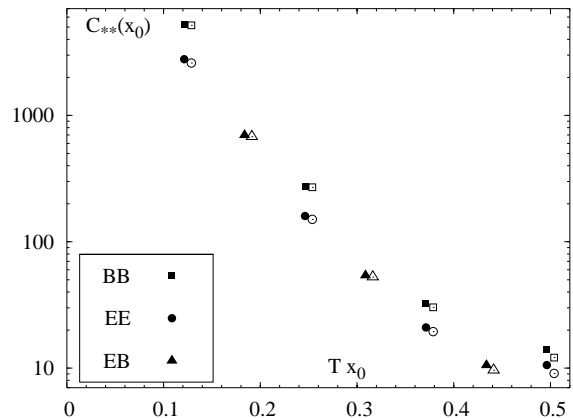,angle=-90,width=8.5cm}
\end{center}
\caption{The correlators that contribute to $C(x_0)= \quarter(C_{BB}+ C_{EE}+2C_{EB})$.
Filled symbols correspond to $T=1.65T_c$, open symbols to $1.24T_c$. Error bars are smaller
than the data symbols.}
\la{fig:Cshear-log}
\end{figure}

The three electric-electric, magnetic-magnetic and electric-magnetic 
contributions to $C(x_0)$ are computed separately and shown on \fig\ref{fig:Cshear-log}.
We apply the following technique to remove the tree-level discretization errors~\cite{sommer} 
separately to $C_{BB},~ C_{EE}$ and $C_{EB}$.
Firstly, $\bar x_0$ is defined such that $C_{\rm cont}^{\rm t.l.}(\bar x_0)=C_{\rm lat}^{\rm t.l.}(x_0)$.
The improved correlator is defined at a discrete set of points through 
$\overline C(\bar x_0) = C( x_0)$, and then augmented to a continuous function via 
$\overline C(\bar x_0^{(i)}) = \alpha + \beta C_{\rm cont}^{\rm t.l.}(\bar x_0^{(i)}) $, $i=1,2$,
where $\bar x_0^{(1)}$ and $\bar x_0^{(2)}$ correspond to two adjacent measurements. 

The resulting improved correlator, normalized by the continuum tree-level result,
is shown on \fig\ref{fig:Cshear-tlimp}. One observes that the deviations from the tree-level result 
are surprisingly small, while deviations from conformality are visible.
The latter is not unexpected at these temperatures, 
where $p/T^4$ is still strongly rising~\cite{Boyd:1996bx}.
Finite-volume effects on the $T=1.65T_c$ lattice are smaller than one part in $10^3$ at tree-level.
Non-perturbatively, at the same temperature with resolution $L_0/a=6$, 
increasing $L/a$ from 20 to 30 reduces $\ovl C(L_0/2)$ by a factor 0.922(73).
While not statistically compelling as it stands, 
the effect deserves further investigation.

\begin{figure}
\begin{center}
\psfig{file=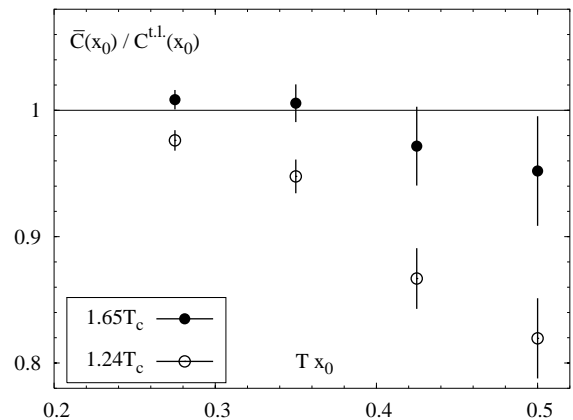,angle=-90,width=8.5cm}
\end{center}
\caption{The tree-level improved correlator $\ovl C(x_0)$ normalized to the tree-level continuum 
         infinite-volume prediction. 
The four points in each sequence are strongly correlated, 
but their covariance matrix is non-singular.}
\la{fig:Cshear-tlimp}
\end{figure}

The entropy density is obtained from the relation $s=(\epsilon+p)/T$ and 
the standard method to compute $\epsilon+p$~(\cite{karsch-aniso}, Eq.1.14). We find 
$s/T^3 = 4.72(3)(5)$ and $ 5.70(2)(6)$ respectively at $T/T_c=1.24$ and $1.65$ 
(the first error is statistical and the second is the uncertainty on $Z(g_0)$).
The Stefan-Boltzmann value is $32\pi^2/45$ in the continuum and $1.0867$ times 
that value~\cite{karsch-aniso} at $L_0/a=8$.

\noindent\textit{Unsatisfactory attempts to extract the viscosity.---~~}
In order to compare with previous studies~\cite{karsch-visco,nakamura}, 
we fit $\ovl C(x_0)$ with a Breit-Wigner ansatz
\be
\rho(\omega)/\omega = \frac{F}{1+b^2(\omega-\omega_0)^2} + \frac{F}{1+b^2(\omega+\omega_0)^2},
\ee
although it clearly ignores asymptotic freedom, 
which implies that $\rho(\omega)\sim \omega^4$ at $\omega \gg T$~\cite{aarts}.
The result of a correlated fit at $T=1.65T_c$ using the points at $Tx_0=0.5$, 0.35 and 0.275
is $a^3F=0.78(4)$, $(b/a)^2=240(30)$ and $a\omega_0=2.36(4)$, and hence
$\eta/s|_{T=1.65T_c}= 0.33(3)$.  A comparison of this to the results of Ref.~\cite{nakamura}
illustrates the progress made in statistical accuracy.

An ansatz motivated by the hard-thermal-loop framework is~\cite{aarts}
\be
\rho(\omega)/\omega = \frac{\eta/\pi}{1+b^2\omega^2} + 
                       \theta(\omega-\omega_1)\frac{A\omega^3}{\tanh\omega/4T}.
\ee
It is capable of reproducing the tree-level prediction, \eq\ref{eq:tl}, and 
it allows for a thermal broadening of the delta function at the origin.
Fitting the $T=1.65T_c$ points shown on \fig\ref{fig:Cshear-tlimp},
the $\chi^2$ is minimized for $b=0$ (effectively eliminating a free parameter),
$A/A_{\rm t.l.}=0.996(8)$, $\omega_1/T=7.5(2)$ and $\eta/s= 0.25(3)$, 
with $\chi^2_{\rm min}=4.0$.
Thus while the ansatz is hardly compatible with the data, it shows
that the data tightly constrains the coefficient $A$ to assume its tree-level value.

\vspace{0.3cm}
\noindent\textit{A bound on the viscosity.---~~}
The positivity property of $\rho(\omega)$ allows us to derive an upper bound on the
viscosity, based on the following assumptions:
\begin{enumerate}
\item the contribution to the correlator from $\omega>\Lambda$ is correctly predicted by 
      the tree-level formula
\item the width of any potential peak in the region $\omega<T$ is no less than O($T$).
\end{enumerate}
The standard QCD sum rule practice is to use perturbation theory from the energy 
lying midway between the lightest state and the first excitation.
With this in mind we choose 
$ \Lambda={\rm max}(\half[M_2+M_{2^*}]\approx2.6{\rm GeV},5T),$
where $M_{2^{(*)}}$ are the masses of
the two lightest tensor glueballs. Perturbation theory predicts a 
Breit-Wigner centered at the origin of width $\Gamma=2\gamma$~\cite{aarts}, 
where $\gamma\approx \alpha_sNT$ 
is the gluon damping rate. To derive the upper bound we conservatively assume that 
for $\omega<\sqrt{2}T$, $\rho(\omega)/\omega$ is a Breit-Wigner  
of width $\Gamma=T$ centered at the origin. From
$ C(\half L_0) \geq L_0^5\left[\int_0^{\sqrt{2}T}\rho_{BW}(\omega) 
            + \int_\Lambda^\infty \rho_{\rm t.l.}(\omega) \right]
             \frac{d\omega}{\sinh\omega L_0/2} $
we obtain (with $90\%$ statistical confidence level)
\be
\eta/s < \left\{ \begin{array}{l@{~~~}l}
 0.96 & (T=1.65T_c) \\
 1.08 & (T=1.24T_c)    . 
          \end{array} \right.
\la{eq:upbo}
\ee

\noindent\textit{The spectral function.---~~}
\begin{figure}
\begin{center}
\psfig{file=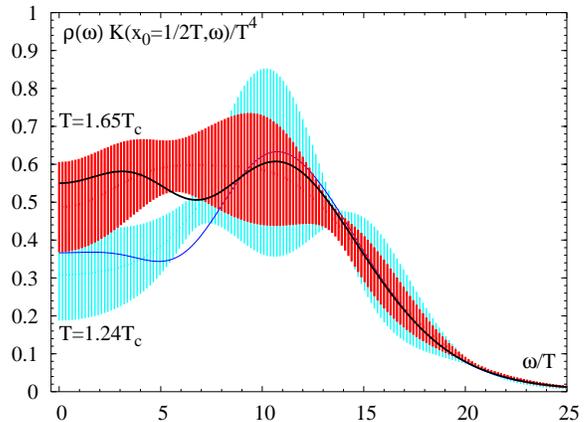,angle=-90,width=8.5cm}
\end{center}
\caption{The result for $\rho(\omega)$. The meaning of the error bands 
and the curves is described in the text.
The area under them equals $\ovl C(L_0/2)= 8.05(31)$ and 9.35(42) for 
$1.24T_c$ and $1.65T_c$ respectively.}
\la{fig:sf}
\end{figure}
As illustrated above, it is rather difficult to find a functional form for $\rho(\omega)$
that is both physically motivated and fits the data.
In a more model-independent approach, $\rho(\omega)$ is expanded in an 
orthogonal set of functions, which grows
as the lattice resolution on the correlator increases, 
and becomes complete in the limit of $L_0/a\to\infty$.
We proceed to determine the function 
$\bar\rho(\omega)\equiv  \rho(\omega)/\tanh(\half\omega L_0)$ by making the ansatz
\be
\bar\rho(\omega) = m(\omega)~[1+a(\omega)], \la{eq:ansatz} 
\ee
where $m(\omega)>0$ has the high-frequency behavior of \eq\ref{eq:tl}, 
and correspondingly define 
$\bar K(x_0,\omega) = \cosh\omega(x_0-\half  L_0)/\cosh\half\omega L_0$.
Suppose that $m(\omega)$ already is a smooth approximate solution to $\bar\rho(\omega)$;
inserting (\ref{eq:ansatz}) into (\eq\ref{eq:C=int_rho}), one requires that 
$a(\omega)=\sum_\ell c_\ell a_\ell(\omega)$, with $\{a_\ell\}$ a basis of functions
which is as sensitive as possible to the discrepancy between the lattice correlator
and the correlator generated by $m(\omega)$. 
These are the eigenfunctions of largest eigenvalue of the 
symmetric kernel
 $G(\omega,\omega')\equiv\int_0^{L_0} \frac{dx_0}{L_0} M(x_0,\omega) M(x_0,\omega')$,
where $M(x_0,\omega)\equiv \bar K(x_0,\omega) m(\omega)$.
These functions  satisfy
$\int_0^\infty d\omega u_\ell(\omega) u_{\ell'}(\omega)=\delta_{\ell\ell'}$ and have an
increasing number of nodes as their eigenvalue decreases.
Thus the more data points available, the larger the basis and the finer details of 
the spectral function one is able to determine.

To determine the spectral function from $N$ points of the correlator, 
we proceed by first discretizing the $\omega$ variable into an $N_\omega$-vector.
The final spectral function is given by the last member $\rho^{(N)}$ of a sequence whose first 
member is $\rho^{(0)}=m$ and whose general member
$\rho^{(n)}$ reproduces $n$ points (or linear combinations) of the lattice correlator.
For $n\geq1$, $\rho^{(n)}=\rho^{(n-1)} [1+\sum_{\ell=1}^n c^{(n)}_\ell a^{(n)}_\ell]$
and the functions $a^{(n)}_\ell(\omega)$ are found by the SVD decomposition~\cite{bryan}
of the $N_\omega\times n$  matrix $M^{(n)t}$, where 
$M_{ij}^{(n)}\equiv\bar K(x_0^{(i)},\omega_j)\bar\rho^{(n-1)}(\omega_j)$.
The `model' $m(\omega)$ is thus updated and agrees with $\rho(\omega)$ at the 
end of the procedure. We first performed this procedure on coarser lattices with $L_0/a=6$
at the same temperatures, starting from 
$m(\omega)= A_{\rm t.l.}\omega^4/(\tanh(\quarter\omega L_0)\tanh(\half\omega L_0)
\tanh^2(c \omega L_0))$ with $\quarter \leq c\leq \half$, and then recycled 
the output as seed for the $L_0/a=8$ lattices. On the latter we used the $N=4$ points 
shown on \fig\ref{fig:Cshear-tlimp}.

The next question to address is the uncertainty on $\rho(\omega)$.
It is important to realize that even in the absence of statistical errors, 
a systematic uncertainty subsists due to the finite number of basis functions 
we can afford to describe $\rho(\omega)$ with. 
A reasonable measure of this 
uncertainty is by how much $\rho(\omega)$ varies if one doubles the resolution on
$C(x_0)$. This can be estimated by `generating' new points 
by using the computed $\rho^{(N)}(\omega)$.
On the other hand we perform a two-point interpolation in $x_0$-space (we chose the form
$(\alpha+\beta(x_0-\half L_0)^2)/\sin^5(\pi x_0/L_0)$), and take 
the difference between these and the generated ones as their systematic uncertainty. 
In practice this difference is added in quadrature 
with the statistical uncertainty. Next we repeat the procedure to find $\rho$
described above with $N \to 2N$: if we use as seed $\rho^{(N)}$, 
then by construction it is left invariant by the iterative procedure, but 
the derivatives of $\rho^{(2N)}$ with respect to the $2N$  points 
of the correlator can be evaluated.
The error on $\rho(\omega)$ is then obtained from a formula of the type
$(\delta\rho)^2=\sum_{i=1}^{2N} (\frac{\partial \rho}{\partial C_i})^2 (\delta C_i)^2$ which
however keeps track of correlations in $x_0$ and Monte-Carlo time.
This is the error band shown on \fig\ref{fig:sf} and the corresponding
shear viscosity values are
\be
\eta/s = \left\{ \begin{array}{l@{~~~}l}
   0.134(33) & (T=1.65T_c) \\
 0.102(56) & (T=1.24T_c).     
          \end{array} \right.
\la{eq:eta}
\ee
It is also interesting to check 
for the stability of the solution under the use of a larger basis of functions. 
If instead of starting from $\rho^{(N)}(\omega)$ we restart from $\rho^{(0)}$
(the output of the $L_0/a=6$ lattice)
and fit the $2N$ (dependent) points using $2N$ basis functions $\{a_\ell\}$,
we obtain the curves drawn on \fig\ref{fig:sf}. As one would hope, 
the oscillations of $\rho^{(2N)}(\omega)$ are covered by the error band.

%
\noindent\textit{Conclusion.---~~}
Using state-of-the-art lattice techniques, we have computed 
the correlation functions of the energy-momentum tensor to high accuracy
in the SU(3) pure gauge theory.
We have calculated the leading high-temperature cutoff effects and removed them
from the correlator relevant to the shear viscosity, and we normalized it
non-perturbatively, exploiting existing results. We obtained the entropy density
with an accuracy of $1\%$.
The most robust result obtained on the shear viscosity is the upper bound 
\eq(\ref{eq:upbo}), which comes from lumping the area under the curve on \fig\ref{fig:sf}
in the interval $[0,6T]$ 
into a peak  of width $\Gamma=T$ centered at the origin.
Secondly, our best estimate of the shear viscosity is given by \eq(\ref{eq:eta}),
using a new method of extraction of the spectral function. 
The errors contain an estimate of the systematic uncertainty associated 
with the limited  resolution in Euclidean time. 
We are extending the calculation to finer lattice spacings and larger 
volumes to further consolidate our findings.

The values (\ref{eq:eta}) are intriguingly close to saturating
the KSS bound~\cite{kovtun} $\eta/s\geq1/4\pi$. 
We note that in perturbation theory
the ratio $\eta/s$ does not depend strongly on the number of 
quark flavors~\cite{arnold}.
Our results thus corroborate the picture of a near-perfect fluid 
that has emerged from the RHIC experiments,
with the magnitude of the anisotropic flow incompatible with 
$\eta/s\gtrsim 0.2$~\cite{teaney}.

\noindent\textit{Acknowledgments.---~~}
I thank Krishna Rajagopal and Philippe de Forcrand for 
their encouragement and many useful discussions. This work was supported in part by 
funds provided by the U.S. Department of Energy under cooperative research agreement
DE-FC02-94ER40818.



\end{document}